\documentclass[aps,prc,preprint,groupedaddress,amsmath,amssymb]{revtex4}

\usepackage{graphicx}
\usepackage{dcolumn}
\usepackage{bm}

\begin{document}


\title{Study of the effect of the tensor correlation on the alpha clustering in $^8$Be with the charge- and parity-projected
Hartree-Fock method}


\author{Satoru Sugimoto}
\email[email: ]{satoru@ruby.scphys.kyoto-u.ac.jp}
\affiliation{Kyoto
University, Kitashirakawa, Kyoto 606-8502, Japan}
\author{Kiyomi Ikeda}
\email[email: ]{k-ikeda@riken.jp}
\affiliation{The Institute of Physical and Chemical Research
(RIKEN), Wako, Saitama 351-0198, Japan}
\author{Hiroshi Toki}
\email[email: ]{toki@rcnp.osaka-u.ac.jp}
\affiliation{Research Center
for Nuclear Physics (RCNP), Osaka University, Ibaraki, Osaka
567-0047, Japan}

\date{\today}

\begin{abstract}
We study the effect of the tensor correlation in the alpha
clustering in $^8$Be. We take as the wave function of the alpha
particle the one calculated by the charge- and parity-projected
Hartree- Fock method, which was proposed by us recently. We use the
wave function in a cluster-model calculation. The expectation value
of the potential energy from the tensor force is almost double of
that in the alpha particle. The energy surface as the function of
the relative distance of two alpha particles becomes much steeper in
an inner region by the inclusion of the tensor correlation in the
wave function of the alpha particle. This is caused by the $p$-state
mixing induced by the tensor correlation. By superposing the wave
function with different relative distances, the reasonable binding
energy of $^8$Be is obtained.
\end{abstract}


\maketitle

\section{Introduction}
The understanding of nuclear structure based on the realistic
nuclear force is one of the main issue of nuclear physics. The
strong tensor force is a characteristic feature of the nuclear force
and is known to play important roles in nuclear structure.
In the clustering structure of nuclei, the tensor force is believed
to be essential. The reaction matrix ($G$-matrix) calculation for
$^8$Be and $^{12}$C adopting molecular orbitals as single-particle
states \cite{bando70,yamamoto78} showed that the starting energies
in the $G$-matrix equation become smaller when an alpha-clustering
structure develops and, as the result, the $G$ matrices in the
triplet-even channel become more attractive. This attraction is
caused mainly by the tensor force and enhances the alpha-clustering
structure. The Argonne-Illinois group performed the variational
Monte Carlo and Green's function Monte Carlo calculations with the
realistic nuclear force systematically in light nuclei. They found
that $^8$Be has a well-developed two-alpha cluster
structure.\cite{wiringa00} In their result a large attraction energy
comes from the one-pion-exchange potential. This result infers the
importance of the tensor force in alpha clustering.

For the alpha clustering in the Be isotopes, there are many studies
using various models.\cite{hiura72,abe73,seya81,enyo95} They shows
the importance of the alpha-clustering structure in the Be isotopes.
However, the tensor force is not usually treated explicitly there.
The effect of the tensor force is included implicitly by
renormalizing the central and LS parts of effective interactions
appropriate to model spaces. Recently, there are attempts that try
to treat the tensor force or the pion explicitly by expanding usual
model spaces.\cite{toki02,akaishi04,sugimoto04,ogawa04,myo05} We
proposed a mean-field-type model (the charge- and parity-projected
Hartree-Fock (CPPHF) method), which can treat the tensor force
explicitly by mixing parities and charge states in single-particle
states.\cite{sugimoto04} We applied the CPPHF method to the alpha
particle and found that the tensor correlation can be treated in our
method. Because two parities are mixed in a single-particle state, a
$p$-state component appears in the single-particle wave function in
addition to an $s$-state component. By performing the parity
projection on the total wave function consisting of the
single-particle states with parity mixing, the wave function with
2-particle--2-hole correlations ($(0s)^2(0p)^2$) and
4-particle--4-hole correlations ($(0p)^4$) is
obtained.\cite{toki02,sugimoto04} The $p$-state component is induced
by the tensor force in the $(0s)^4$ configuration, which is usually
assumed as a wave function of the alpha particle. The $p$-state
component corresponds to the $D$-state probability and is not
treated in usual model calculations explicitly. In fact, in the
$G$-matrix calculation in Refs.~\cite{bando70,yamamoto78}, the
effect of the $p$-state component is included in the $G$ matrix as
the effective interaction in the model space which essentially
consists of the $(0s)^4$ configuration. Therefore it is interesting
to study $^8$Be based on the CPPHF method.

In the present paper, we make an alpha-cluster model using the wave
function of the alpha particle calculated by the CPPHF method and
apply it to $^8$Be to see the effect of the tensor force on the
alpha clustering. In Section~\ref{sec:formulation} we formulate the
alpha-cluster model with the wave function of the alpha particle
calculated in the CPPHF method and in Section~\ref{sec:result} we
apply it to $^8$Be. In Section~\ref{sec:summary} we summarize the
paper.
\section{Formulation}\label{sec:formulation}
In the present study, the wave function of the alpha particle is
calculated with the charge- and parity-projected Hartree-Fock
(CPPHF) method.\cite{sugimoto04} In the CPPHF method, the wave
function of the alpha particle has the following form,
\begin{align}
\Psi_\alpha=P^\text{C}(Z) P^\text{P}(\pm) \Phi_\alpha .
\label{eq:alpha}
\end{align}
Here, $\Phi_\alpha$ is a Slater determinant composed of
single-particle states with charge and parity mixing and, therefore,
does not have a good parity and a definite charge number. To exploit
a wave function having a good parity, positive ($+$) or negative
($-$), and a definite charge number $Z$, the parity-projection
operator $P^\text{P}(\pm)$ and the charge-projection operator
$P^\text{C}(Z)$ are operated on $\Phi_\alpha$. In the present study,
the wave function of the alpha particle is fixed to the ground state
and, then, the parity is positive and $Z$ is equal to 2. The detail
of the CPPHF method is found in Ref.~\onlinecite{sugimoto04}.
$\Phi_\alpha$ can be thought as a kind of an intrinsic wave
function. Assuming the spherical symmetry, the intrinsic wave
function can be written as
\begin{align}
\label{eq:alpha_int}
\Phi_\alpha=\mathcal{A} \prod_{i=1}^4 \Bigl(
\phi_{s_{1/2};\pi} (r_i) \mathcal{Y}_{0 \frac{1}{2} m_i} (\Omega_i)
\zeta_{\frac{1}{2}\frac{1}{2}}(i)+\phi_{s_{1/2};\nu} (r_i)
\mathcal{Y}_{0 \frac{1}{2} m_i} (\Omega_i)
\zeta_{\frac{1}{2}-\frac{1}{2}}(i)\\ \notag +\phi_{p_{1/2};\pi}
(r_i) \mathcal{Y}_{1 \frac{1}{2} m_i} (\Omega_i)
\zeta_{\frac{1}{2}\frac{1}{2}}(i)+\phi_{p_{1/2};\nu} (r_i)
\mathcal{Y}_{1 \frac{1}{2} m_i} (\Omega_i)
\zeta_{\frac{1}{2}-\frac{1}{2}}(i) \Bigr).
\end{align}
In the above equation, $\phi_{j;\sigma}$ is a radial wave function
for the component with the total angular momentum $j$ ($\sigma=\pi$
for proton and $\sigma=\nu$ for neutron), $\mathcal{Y}_{ljm}$ is an
eigenfunction of the total spin
$\boldsymbol{j}=\boldsymbol{l}+\boldsymbol{s}$, and $\zeta_{1/2m_t}$
is an isospin wave function for proton when $m_t=1/2$ or for neutron
when $m_t=-1/2$. $\Phi_\alpha$ has $p$-state components, which are
induced by the tensor force. By performing the parity and charge
projections on the intrinsic wave function with parity and charge
mixing as in Eq.~(\ref{eq:alpha}), the correlated wave function
which have 2-particle--2-hole ($(os)^2(0p)^2$) and
4-particle--4-hole ($(0p)^4$) components is
obtained.\cite{toki02,sugimoto04} We should note that the widths of
the $p$-state components are narrower compared to those of the
$s$-state ones.\cite{akaishi04,sugimoto04,myo05} It means that to
gain the energy from the tensor force we need to treat high-momentum
components, which are not included in a usual mean-field or a
shell-model calculation.

As a wave function of $^8$Be, we put two alpha particles which have
a finite relative distance. Actually $^8$Be is not bound but we
treat it in the bound-state approximation by fixing the relative
distance. By putting the wave functions of the alpha particle along
the z axis, the wave function of $^8$Be becomes
\begin{align}
\Psi_{^8\text{Be}}(R)=N(R)
\mathcal{A}_{1-8}\left[\Psi_\alpha(1,2,3,4; R/2)
\Psi_\alpha(5,6,7,8;-R/2) \right].
\end{align}
Here, the integer numbers from 1 to 8 label nucleons before the
antisymmetrization, $R$ is the relative distance between the two
alpha particles, and $N(R)$ is a normalization factor.
$\Psi_\alpha(i,j,k,l;d)$ is the wave function of the alpha particle
which consists of the nucleons having particle numbers $i$, $j$,
$k$, and $l$ and located at $z = d$. $\Psi_\alpha(i,j,k,l;d)$ is
performed by the parity and charge projections and, therefore, have
a good parity (+) and a definite charge number ($Z=2$) as in
Eq.~(\ref{eq:alpha}). It is important to perform the charge and
parity projections on the intrinsic wave functions located at $z =
R/2$ and $-R/2$ respectively. By doing so, the wave function
$\Psi_{^8\text{Be}}(R)$ is going into the wave functions of two
isolated alpha particles when $R \rightarrow \infty$.
$\mathcal{A}_{1-8}$ is the antisymmetrization operator for all 8
particles.

The wave function $\Psi_{^8\text{Be}}(R)$ is not spherical symmetric
but deformed axially symmetrically. Then we need to perform the
angular momentum projection to obtain a wave function which has a
good angular momentum. In principle we can do such a calculation,
but it is too time consuming. Therefore we only subtract the
expectation value of rotational energy in the rigid rotor
approximation, $\Delta E_\text{rot}$, as in
Ref.~\onlinecite{seya81}. $\Delta E_\text{ROT}$ is defined as
\begin{align}
\Delta E_\text{ROT} = \frac{\hbar^2}{2 I} \left<
\Psi_{^8\text{Be}}\right| \boldsymbol{J}^2 \left|\Psi_{^8\text{Be}}
\right> \label{eq:Erot}
\end{align}
with the momentum of inertia around the $y$ axis,
\begin{align}
I = M \left< \Psi_{^8\text{Be}}\right|
\sum_{i=1}^8(x_i-X_\text{G})^2+ (z_i-Z_\text{G})^2\left|
\Psi_{^8\text{Be}}\right> .
\end{align}
Here, $M$ is the mass of nucleon, $X_\text{G}$ and $Z_\text{G}$ are
the $x$ and $z$ components of the coordinate of the center of mass
of all nucleons, and $\boldsymbol{J}$ is the total angular momentum
$\sum_{i=1}^8 (\boldsymbol{l}_i+\boldsymbol{s}_i)$.

For the Hamiltonian, we take the same form as in
Ref.~\onlinecite{sugimoto04},
\begin{align} H =
-\sum_{i=1}^8\frac{\hbar^2}{2 M} \triangle_i+\sum_{1 \le i < j \le
8} (V^\text{C}_{ij}+V^\text{T}_{ij})-E_\text{G} .
\label{eq:Hamiltonian}
\end{align}
Here, $E_\text{G}$ is the energy of the center of mass motion,
$V^\text{C}$ is the potential energy from the central force, and
$V^\text{T}$ is the potential energy from the tensor force. We use
the Volkov No.~1 force \cite{volkov65} for the central part
($V_\text{C}$) and the G3RS force \cite{tamagaki68} for the tensor
part ($V_\text{T}$). For simplicity, we omit the LS and Coulomb
forces in the present study. We expect that the inclusion of these
forces does not change the results so much.

The effect of the tensor force is already included in the Volkov
No.~1 force, because the binding energy of the alpha particle can be
reproduced in the absence of the tensor force with the Volkov No.~1
force. Because the effect of the tensor force is thought to appear
as attraction in the triplet even part of the central force, we
reduce the attractive part of the Volkov No.~1 force in the triplet
even channel by multiplying a numerical factor $x_\text{TE}$.

In the present study we assume the intrinsic wave function of the
alpha particle as spherical symmetric and, then, the CPPHF method
can take into account only the coupling between $s_{1/2}$ and
$p_{1/2}$ by the tensor force. However, the couplings of $s_{1/2}$
to $p_{3/2}$, $d_{3/2}$, and $f_{5/2}$ are found to be also
important to gain the attractive energy from the tensor force in the
study of the shell-model-type calculation proposed by Myo et
al.\cite{myo05} By considering this, we change the strength of the
tensor force by multiplying the $\vec{\tau}_1\cdot\vec{\tau}_2$ part
of the tensor force by a numerical factor $x_\text{T}$, which is a
dominant part in the tensor force.
\section{Result}\label{sec:result}
\begin{table}[htb]
\caption{\label{tbl:alpha} Results for the ground state properties
of the alpha particle in the simple Hartree-Fock (HF), the
parity-projected Hartree-Fock (PPHF), and the charge- and
parity-projected Hartree-Fock (CPPHF) schemes. $x_\text{T}$ and
$x_\text{TE}$ are the numerical factors multiplied to the
$\vec{\tau}_1\cdot\vec{\tau}_2$ part of the tensor force and the
attractive part of the central force in the triplet-even channel
respectively. $E_\text{TOT}$, $K_\text{TOT}$, $V_\text{TOT}$,
$V_\text{C}$, and $V_\text{T}$ are the expectation values for the
total energy, the kinetic energy, the sum of the potential energies,
the potential energy from the central force, and the potential
energy from the tensor force, which are given in MeV. $R_\text{m}$
is the matter radius in fm and $P(D)$ is the $D$-state probability
in \%.}
\begin{ruledtabular}
\begin{tabular}{crrrrrrrrr}
    & $x_\text{T}$ & $x_\text{TE}$ & $E_\text{TOT}$ & $K_\text{TOT}$
    & $V_\text{TOT}$ & $V_\text{C}$ & $V_\text{T}$ & $R_\text{m}$ & $P(D)$\\
    \hline
    HF & 0.00 & 1.00 & -27.92 & 49.72 & -77.64 & -77.64 & 0.00 & 1.45 & 0.00 \\
  PPHF & 1.00 & 0.96 & -28.26 & 52.50 & -80.76 & -75.07 & -5.68 & 1.45 & 0.68 \\
  PPHF & 1.50 & 0.91 & -28.60 & 57.31 & -85.92 & -72.07 & -13.84 & 1.42 & 1.76 \\
  CPPHF & 1.00 & 0.92 & -28.74 & 57.80 & -86.54 & -73.23 & -13.31 & 1.42 & 3.22 \\
  CPPHF & 1.50 & 0.79 & -28.58 & 70.27 & -98.86 & -63.58 & -35.28 & 1.36 & 8.65
\end{tabular}
\end{ruledtabular}
\end{table}
In the Table~\ref{tbl:alpha}, the results for the properties of the
alpha particle in the simple Hartree-Fock (HF), the parity-projected
Hartree-Fock (PPHF), and the charge- and parity-projected
Hartree-Fock schemes (CPPHF) are summarized. In the PPHF scheme,
only the parity projection is performed. We take two values for
$x_\text{T}$, 1.00 (the normal tensor force case) and 1.50 (the
strong tensor force case). For each $x_\text{T}$, the value of
$x_\text{TE}$ is determined to reproduce the binding energy of the
alpha particle. In the HF case, the single-particle wave function is
fixed to the $0s$ harmonic oscillator wave function with the
oscillator length $b=1.37$ fm. Although we set $x_\text{T}$ to 0.00
in the simple HF calculation, the potential energy from the tensor
force ($V_\text{T}$) becomes zero even if we use the finite value
for $x_\text{T}$, because the wave function is the simple $(0s)^4$
configuration.  From the table, you can see that the charge
projection is effective to gain the energy from the tensor force. By
performing the charge projection in addition to the parity
projection, $V_\text{T}$ becomes almost three-time larger. $P(D)$ is
the $D$-state probability, which is defined as the probability of
the component with the total spin $S=2$.\cite{sugimoto04}

\begin{table}[htb]
\caption{\label{tbl:be8} Results for $^8$Be with the relative
distance $R$ = 3 fm. $x_\text{T}$, $x_\text{TE}$, $E_\text{TOT}$,
$K_\text{TOT}$, $V_\text{TOT}$, $V_\text{C}$, $V_\text{T}$, and
$R_\text{m}$ have the same meanings as in the Table~\ref{tbl:alpha}.
$\Delta E_\text{ROT}$ is the rotational energy defined in
Eq.~(\ref{eq:Erot}) and $E_\text{TOT} =
K_\text{TOT}+V_\text{TOT}-\Delta E_\text{ROT}$.}
\begin{ruledtabular}
\begin{tabular}{crrrrrrrrr}
   & $x_\text{T}$ & $x_\text{TE}$ & $E_\text{TOT}$ & $K_\text{TOT}$
    & $V_\text{TOT}$ & $V_\text{C}$ & $V_\text{T}$ &$\Delta E_\text{ROT}$ & $R_\text{m}$\\
    \hline
HF  & 0.00 &   1.00  &  -52.84 & 126.65 & -172.24 & -172.24 & 0.00 &
7.25 & 2.22\\
PPHF &   1.00 &   0.96 &   -52.50 &  134.48 & -179.15 & -166.66 &
-12.48 &  7.83  &  2.21\\
PPHF  &  1.50  &  0.91  &  -52.12 & 145.95  & -189.41 & -159.47 &
-29.94 & 8.66  &  2.18\\
CPPHF  & 1.00  & 0.92 & -52.90  & 146.90 & -190.96 & -162.05  &
-28.91
& 8.83 & 2.18\\
CPPHF & 1.50 & 0.79  & -50.03 & 175.65  & -214.66  & -139.94 &-74.72
& 11.02 & 2.12
\end{tabular}
\end{ruledtabular}
\end{table}
In Table~\ref{tbl:be8} we show the results for $^8$Be. The relative
distance $R$ is fixed to 3 fm. The energy gains from the correlation
between alpha particles are around 20 MeV for the central force and
several MeV for the tensor force. The increases of the kinetic
energy are around 30 MeV. The rotational energies defined in
Eq.~(\ref{eq:Erot}) are about 10 MeV. The increase of the rotational
energy for the cases with the strong tensor force, is caused by the
two factors. One is the increase of the expectation value of
$\boldsymbol{J}^2$ due to the mixing of $p$ state in the wave
function of the alpha particle and the other is the decrease of the
momentum of inertia $I$ due to the shrinkage of the radius of the
alpha particle. Both are caused by the tensor force.

\begin{figure}[htb]
\includegraphics{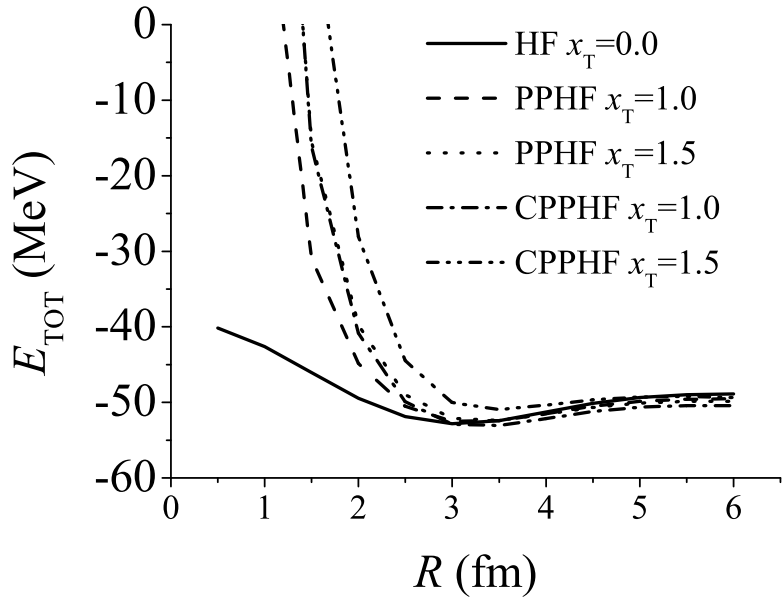}
\caption{\label{fig:ETOT} The total energies $E_\text{TOT}$ of
$^8$Be as the functions of the relative distance $R$ in various
schemes. The solid line is the result for the simple Hartree-Fock
(HF) scheme without the tensor force, the dashed line for the
parity-projected Hartree-Fock (PPHF) scheme with $x_\text{T}$ = 1.0,
the dotted line for the PPHF scheme with $x_\text{T}$ = 1.5, the
dashed-and-dotted line for the charge- and parity-projected
Hartree-Fock (CPPHF) scheme with $x_\text{T}$ = 1.0, and the
dash-and-double-dotted line for the CPPHF scheme with $x_\text{T}$ =
1.5.}
\end{figure}
To see the relative-distance dependence of the total energy, we show
the energy surfaces of $^8$Be as the functions of the relative
distance $R$ in various schemes in Fig.~\ref{fig:ETOT}. In the
figure, the solid line is the result for the simple HF scheme
without the tensor force, the dashed line for the PPHF scheme with
the normal tensor force ($x_\text{T}$ = 1.0), the dotted line for
the PPHF scheme with the strong tensor force ($x_\text{T}$ = 1.5),
the dashed-and-dotted line for the CPPHF scheme with the normal
tensor force ($x_\text{T}$ = 1.0), and the dashed-and-double-dotted
line for the CPPHF scheme with the strong tensor force ($x_\text{T}$
= 1.5). The energy minima appear around $R$ = 3 fm for the simple
Hartree-Fock case and around $R$ = 3.5 fm for the PPHF and CPPHF
cases. The minimum values are between 52 $\sim$ 53 MeV except for
the CPPHF scheme with the strong tensor force. For the CPPHF case
with the strong tensor force the energy minimum is around 51 MeV. In
comparison with the simple HF case without the tensor force, the
energy surfaces become shallower by including the tensor correlation
in the wave functions of the alpha particle. A significant effect of
the tensor correlation appears in an inner region of the energy
surfaces. Here, the energy surfaces for both the PPHF and the CPPHF
cases rise sharply. In fact, the total energy becomes near 300 MeV
at $R$ = 0.5 fm for the CPPHF scheme with the strong tensor force.
In the simple HF case, where there is no tensor correlation, such a
sharp rise of the energy surface does not show up.

\begin{figure}[htb]
\includegraphics{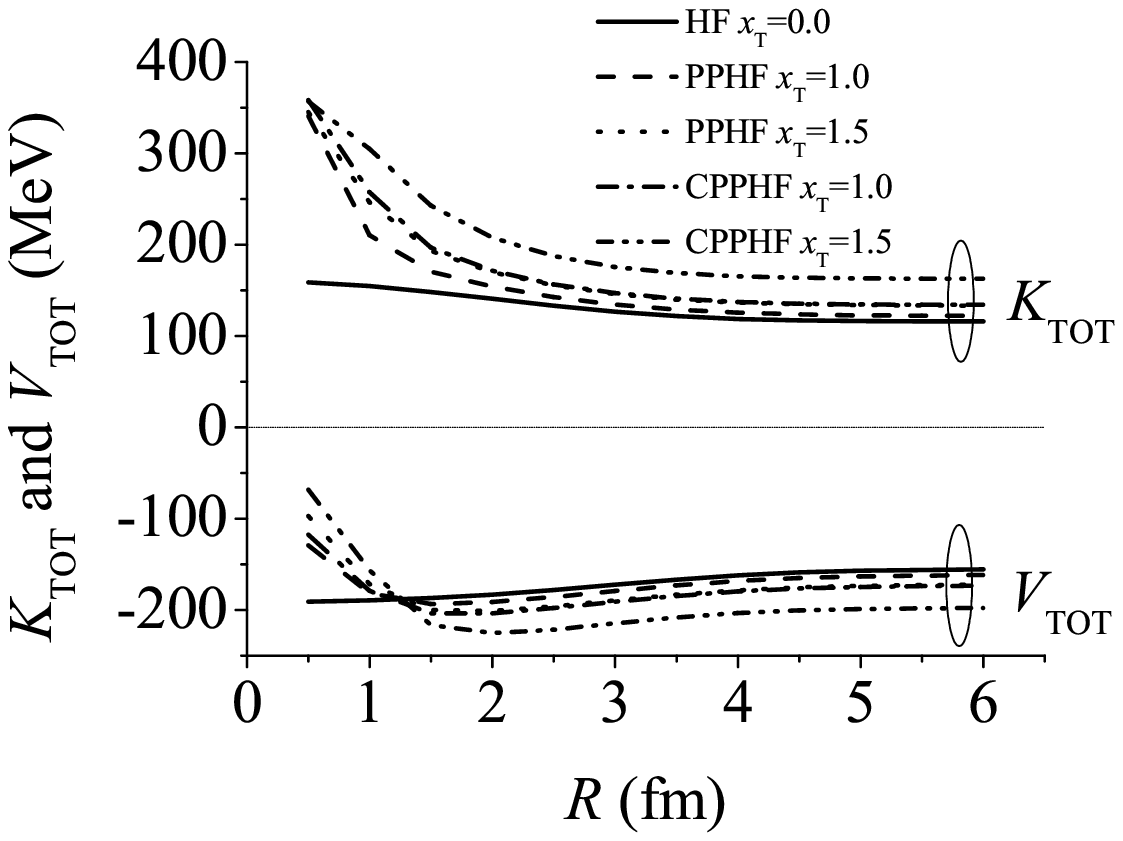}
\caption{\label{fig:KandVTOT} The total kinetic energies
$K_\text{TOT}$ and potential energies $V_\text{TOT}$ of $^8$Be as
the functions of the relative distance $R$ in various schemes. The
meaning of each line is the same as in Fig.~\ref{fig:ETOT}.}
\end{figure}
To check the cause of the sharp increase of the total energy, in
Fig.~\ref{fig:KandVTOT} the contribution to the total energy from
the kinetic energy $K_\text{TOT}$ and the potential energy
$V_\text{TOT}$ are shown separately. The potential energies for all
cases become smaller when the relative distance becomes smaller.
This tendency continues beyond $R$ = 2 fm. In contrast to the
potential energy, the kinetic energy increases monotonically. This
increase of the kinetic energy is more significantly for the cases
with the tensor correlation than for that without the tensor
correlation. These facts indicate that the sharp increase of the
total energy is caused by the steep rise of the kinetic energy. In
the PPHF and CPPHF case, $p$-state components are induced in the
wave functions of the alpha particle by the tensor correlation as
indicated in Eq.~(\ref{eq:alpha_int}). This $p$-state mixing results
in finite $P(D)$ value as shown in Table~\ref{tbl:alpha}. In the
simple HF case, the wave function of the alpha particle is in the
simple $(0s)^4$ configuration and has no $p$-state component.
Furthermore, the $p$-state component induced by the tensor force is
compact in size.\cite{sugimoto04} It implies that the $p$-state
component has high-momentum component and large kinetic energy.
Therefore, the sharp increase of the kinetic energy in the inner
region is thought to be caused by the $p$-state mixing induced by
the tensor force. The repulsion between two alpha particles in the
small distance is usually thought to come mainly from the effect of
the antisymmetrization.\cite{tamagaki65} In the simple HF case the
repulsion in a small distance is mainly due to the effect of the
antisymmetrization and the wave function becomes a deformed-shell
configuration when $R\rightarrow 0$. Our results for the CPPHF and
PPHF cases indicate the correlation by the tensor force is more
important for the repulsion.

\begin{figure}[htb]
\includegraphics{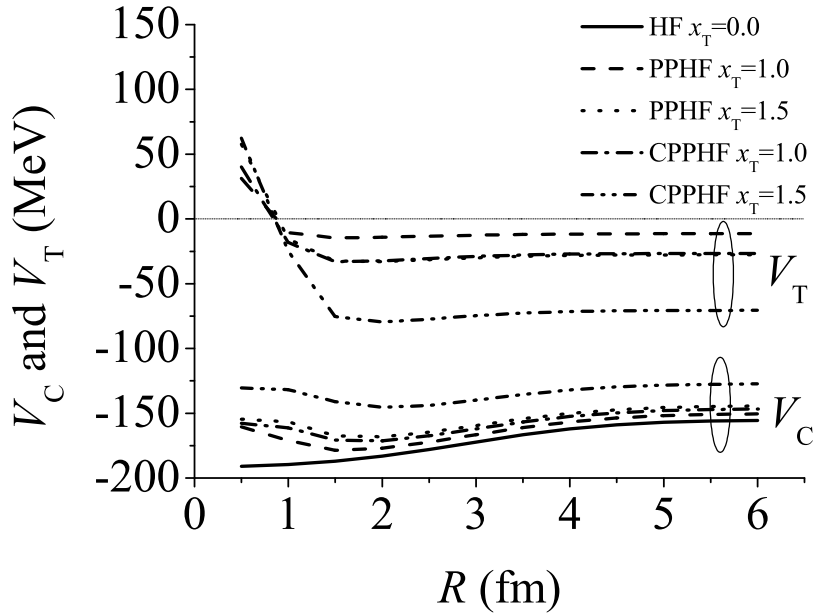}
\caption{\label{fig:VCandVT} The potential energies from the central
force $V_\text{C}$ and the tensor force $V_\text{T}$ of $^8$Be as
the functions of the relative distance $R$ in various schemes. The
meaning of each line is the same as in Fig.~\ref{fig:ETOT}.}
\end{figure}
In Fig.~\ref{fig:VCandVT} we show the energy contribution to the
potential energy from the central force $V_\text{C}$ and the tensor
force $V_\text{T}$ separately. Both the central potential energy and
the tensor potential energy become smaller when the relative
distance of the alpha particle, $R$, becomes smaller for $R \gtrsim$
2 fm. The decrease of the tensor potential energy is less
significant than that of the central potential energy.
When $R$ goes down less than about 1.5 fm the
potential energies become larger. In fact the tensor potential
energy becomes positive for very small $R$. It should be noted that
for the small-relative-distance region the approximation of
two-alpha clusters for the wave function of $^8$Be would not be
valid. The other configurations like deformed alpha clusters and a
deformed configuration of $^8$Be as a whole become important. Such
configurations are not treated here. The inclusion of these
configurations may change energy surface in the inner region.
\begin{table}[htb]
\caption{\label{tbl:gcm} Results of the generator coordinate method
(GCM) calculation using the wave functions with the relative
distances from 1 fm to 11 fm with the equal separation.
$x_\text{T}$, $x_\text{TE}$, $E_\text{TOT}$, $K_\text{TOT}$,
$V_\text{TOT}$, $V_\text{C}$, $V_\text{T}$, $R_\text{m}$, and
$\Delta E_\text{ROT}$ have the same meanings as in the
Table~\ref{tbl:be8}.}
\begin{ruledtabular}
\begin{tabular}{crrrrrrrrr}
 & $x_\text{T}$ & $x_\text{TE}$ & $E_\text{TOT}$ & $K_\text{TOT}$
    & $V_\text{TOT}$ & $V_\text{C}$ & $V_\text{T}$ &$\Delta E_\text{ROT}$ & $R_\text{m}$\\
    \hline
HF & 0.00 & 1.00 & -55.25 & 121.65 & -169.05 & -169.05 & 0.00 & 7.85 & 2.46 \\
PPHF & 1.00 & 0.96 & -56.50 & 123.15 & -170.79 & -159.53 & -11.26 & 8.86 & 2.71 \\
PPHF & 1.50 & 0.91 & -57.93 & 130.76 & -178.56 & -151.62 & -26.94 & 10.13 & 2.89 \\
CPPHF & 1.00 & 0.92 & -58.54 & 131.71 & -180.00 & -154.17 & -25.82 & 10.25 & 2.86 \\
CPPHF & 1.50 & 0.79 & -59.81 & 153.71 & -200.00 & -132.50 & -67.50 &
13.52 & 3.24
\end{tabular}
\end{ruledtabular}
\end{table} %

Finally, we apply the generator coordinate method (GCM)
\cite{hill53} for the relative distance $R$. The wave functions of
several discrete relative distances $R_i$ are superposed as
following,
\begin{align}
\Psi_{^8\text{Be}}^\text{GCM} = \sum_i c_i \Psi_{^8\text{Be}} (R_i)
.
\end{align}
The coefficients $c_i$ are determined by solving the Hill-Wheeler
equation for the discrete $R_i$,
\begin{align}
\sum_j \langle \Psi_{^8\text{Be}} (R_i) |H| \Psi_{^8\text{Be}} (R_j)
 \rangle c_j = E^\text{GCM} \sum_j \langle \Psi_{^8\text{Be}} (R_i) | \Psi_{^8\text{Be}}
(R_j)\rangle c_j .
\end{align}
We subtract from $E^\text{GCM}$ the rotational energy $\Delta
E_\text{ROT}$ in Eq.~(\ref{eq:Erot}) to obtain $E_\text{TOT}$. We
adopt as $R_i$ the six points from 1 fm to 11 fm with the equal
separation. We show the results for the various cases in
Table~\ref{tbl:gcm}. By superposing the several wave functions with
the different relative distance, the reasonable binding energies are
obtained for the cases with the tensor correlation. The energy gains
are larger for the strong tensor force cases. It is due to the
larger expectation value of $\boldsymbol{J}^2$ in $\Delta
E_\text{ROT}$ for the strong tensor force cases. The radii are also
larger for the strong tensor force cases due to the shallower energy
surfaces. We should note that in this work the expectation value of
the rotational energy is subtracted after the variation but the
angular momentum projection may change the results.
\section{Summary}\label{sec:summary}
We formulate a cluster model with the charge- and parity-projected
Hartree-Fock (CPPHF) method to study the effect of the tensor force
on the alpha clustering in nuclei. The wave function of the alpha
particle is calculated by the CPPHF method and used in the
cluster-model calculation. The wave function of the alpha particle
has $p$-state mixing induced by the tensor correlation in addition
to the simple $(0s)^4$ configuration, which are usually assumed in
cluster-model calculations.

We apply the model to $^8$Be. The correlation energy from the tensor
force becomes almost double of the isolated alpha particle. The
kinetic energy becomes much larger for the case with the tensor
correlation due to the $p$-state mixing in the wave function of the
alpha particle. The energy surface becomes much steeper in a small
relative distance by the inclusion of the tensor correlation. The
sharp rise of the energy surfaces is mainly caused by the large
increase of the kinetic energy. It is also due to the $p$-state
mixing in the wave function of the alpha particle. In our result its
effect is stronger than the effect from the antisymmetrization. The
dependence of the potential energy from the tensor force is smaller
than that from the central force. By superposing the wave function
with the different relative distances, the reasonable binding
energies are obtained with the cases with the tensor correlation.

In the present work we only include the $p$-state mixing in the wave
function of the alpha particle. The mixing of higher
angular-momentum components like $d$-state and $f$-state are
probably important. The effect of the dissolution of the alpha
particle in the small relative distance region is also important. It
is not included in the present calculation. Solving the scattering
problem using the wave function of the alpha particle with the
tensor correlation is also interesting. The studies of these are
under progress.
\begin{acknowledgments}
We acknowledge fruitful discussions with Prof. H.~Horiuchi on the
role of the tensor force in light nuclei. A part of the present
study is partially supported by the Grant-in-Aid from the Japan
Society for the Promotion of Science (14340076). This work is
supported by the Grant-in-Aid for the 21st Century COE ``Center for
Diversity and Universality in Physics'' from the Ministry of
Education, Culture, Sports, Science and Technology (MEXT) of Japan.
This work is partially performed in the Research Project for the
Study of Unstable Nuclei from Nuclear Cluster Aspects sponsored by
the Institute of Physical and Chemical Research (RIKEN). A part of
the calculation of the present study was performed on the RIKEN
Super Combined Cluster System (RSCC) in the Institute of Physical
and Chemical Research (RIKEN).
\end{acknowledgments}


\end{document}